\documentclass[prl,twocolumn]{revtex4}
\usepackage{graphicx}
\renewcommand{\v}[1]{{\bf #1}}
\newcommand{\sign}{{\rm sign}}

\newcommand{\w}{{\omega}}

\def\eqa{\begin{eqnarray}}
\def\eea{\end{eqnarray}}
\newcommand{\eq}{\begin{equation}}
\newcommand{\ee}{\end{equation}}
\newcommand{\nn}{\nonumber\\}

\newcommand{\Tr}{{\rm Tr}}

\renewcommand{\Im}{{\rm Im}}

\newcommand{\ra}{\rightarrow}

\newcommand{\al}{\alpha}
\newcommand{\bt}{\beta}
\newcommand{\del}{\delta}
\newcommand{\Del}{\Delta}

\newcommand{\Ga}{\Gamma}

\newcommand{\la}{\lambda}

\newcommand{\si}{\sigma}

\newcommand{\cG}{ {\cal G} }

\begin{document}

\title{Electronic structure near an impurity and terrace on the surface of a 3-dimensional topological insulator}

\author{Qiang-Hua Wang$^1$}
\email[E-mail address:]{qhwang@nju.edu.cn}

\author{Da Wang$^1$}

\author{Fu-Chun Zhang$^2$}

\affiliation{$^1$National Laboratory of Solid State
Microstructures and Department of Physics, Nanjing University,
Nanjing 210093, China}

\affiliation{$^2$Department of Physics, The University of Hong
Kong, Pokfulam Road, Hong Kong, China}

\begin{abstract}

Motivated by recent scanning tunneling microscopy experiments on
surfaces of Bi$_{1-x}$Sb$_{x'}$\cite{yazdanistm,gomesstm} and
Bi$_2$Te$_3$,\cite{kaptunikstm,xuestm} we theoretically study the
electronic structure of a 3-dimensional (3D) topological insulator
in the presence of a local impurity or a domain wall on its
surface using a 3D lattice model. While the local density of
states (LDOS) oscillates significantly in space at energies above
the bulk gap, the oscillation due to the in-gap surface Dirac
fermions are very weak. The extracted modulation wave number as a
function of energy satisfies the Dirac dispersion for in-gap
energies and follows the border of the bulk continuum above the
bulk gap. We have also examined analytically the effects of the
defects by using a pure Dirac fermion model for the surface states
and found that the LDOS decays asymptotically faster at least by a
factor of 1/r than that in normal metals, consistent with the
results obtained from our lattice model.
\end{abstract}

\maketitle

\section{I. Introduction}

A three-dimensional (3D) topological insulator (TI) is a
time-reversal invariant system with bulk gap but support massless
Dirac fermions with coupled spin and momentum on the
surface.\cite{fu-kane,qi-prl} The existence and the (odd) number
of Dirac cones of the massless dispersion are protected by the
$Z_2$ topology of the bulk band structure.\cite{z2} Recent first
principle calculation reveals that Sb$_2$Te$_3$, Bi$_2$Te$_3$ and
Bi$_2$Se$_3$ crystals are potential topological
insulators.\cite{zhang-nphys} Quite excitingly angle-resolved
photoemission (ARPES) experiments have confirmed many of the
proposed topological insulators.\cite{arpes} Before harvesting the
novel properties of TI's,\cite{response} an interesting question
is how robust the surface Dirac fermions are against
imperfections, such as local and extended impurities. This issue
is made interesting and challenging by recent scanning tunnelling
microscopy (STM)
measurements.\cite{yazdanistm,kaptunikstm,xuestm,gomesstm} It is
observed that while local impurities are seen to induce local
density of states (LDOS) oscillation,\cite{xuestm} surprisingly
the domain wall (a terrace on the surface) does not seem to induce
LDOS oscillations for energies within the bulk
gap.\cite{yazdanistm,kaptunikstm,gomesstm}

We investigate the above issue by a model lattice hamiltonian for
a 3-dimensional TI. The main results are as follows. 1) We obtain
the surface Green's function and therefore the spectral function
for the surface states. The contributions from surface Dirac
fermions and the bulk extended states are clearly identified. 2)
For an upper terrace on the surface, we find the LDOS oscillation
is vanishingly weak for energies below the bulk gap $\Del$, where
Dirac fermions are best defined. The oscillation is significant
above $\Del$, but it can be clearly ascribed to the contributions
from the bulk extended states near the bottom of the conduction
band. The numerical phenomenology is in nice agreement with the
experiment.\cite{yazdanistm,kaptunikstm,gomesstm} On the other
hand, the LDOS oscillation on the lower terrace is globally weak.
The asymmetry of the terraces can be attributed the the difference
in the effective scatting mechanism. 3) For a local unitary
impurity on a flat surface, we find LDOS oscillations at all
energies, although it is relatively weaker below $\Del$. 4)
Combining both terraces and local impurities, we extract from the
LDOS oscillations the energy $\w$ dependence of the modulation
wave number $2k_\w$. For $\w<\Del$ the dispersion ($\w$ vs.
$k_\w$) coincides with that of surface Dirac fermions, while for
$\w>\Del$ it is actually related to the bulk extended states. The
above numerical results combine to support the robustness of
surface Dirac fermions. 5) For comparison, we substantiate
analytical results using pure Dirac fermion models subject to
impurity scattering. Asymptotically, the oscillation in LDOS, if
present at all, has an energy dependent wave number $2k_\w$, where
$k_\w$ is the on-shell momentum, and decays faster by a factor of
$1/r$ than that for usual fermions. In particular, a hard domain
wall in 2D Dirac models does not lead to any LDOS oscillations at
all. These are in qualitative agreement with the numerical results
for the full 3d model. However, we do find and discuss differences
between surface and pure Dirac fermions in connection to the
nature of the wave functions.

The rest of the paper is organized as follows. We discuss the
surface states using a 3D lattice model in sections II-IV, where a
perfect surface, a terrace and a local impurity are discussed,
respectively. Sec.V contains analytical results using pure Dirac
models. We summarize and provide remarks in connection to
experiments in Sec.VI.

\section{II. Surface states in a lattice model of topological insulators}

We start with a 3D lattice model for topological insulators. The
hamiltonian is given by $H=\sum_\v k \psi_\v k^\dagger h_\v
k\psi_\v k$, where $\psi_\v k$ is a 4-spinor, $\v k$ is the
lattice momentum, and \eqa h_\v k=[m-\sum_b 2t_b(\cos
k_b-1)]\Ga_0+\sum_b 2t_b\sin k_b \Ga_b,\eea where $m$ is a
parameter controlling the bulk gap, $t_b$ is the hopping amplitude
along a bond $\v b$, $k_b=\v k\cdot \v b$, and
$\Ga_b=(\Ga_1\hat{x}+\Ga_2\hat{y}+\Ga_3\hat{z})\cdot\hat{b}$. Here
$\hat{x}$, $\hat{y}$ and $\hat{z}$ are three orthogonal unit
vectors, $\hat{b}$ is the unit vector along $\v b$, and
$\Ga_\mu$'s (for $\mu=0,1,2,3$) are Hermitian Dirac matrices that
satisfy the Clifford algebra $\{\Ga_\mu,\Ga_\nu\}=2\del_{\mu\nu}$.
Explicitly we take $\Ga_1=\si_1\otimes\tau_3$,
$\Ga_2=\si_2\otimes\tau_3$, $\Ga_3=\si_0\otimes\tau_2$ and
$\Ga_0=\si_0\otimes \tau_1$, where $\si$ and $\tau$ are Pauli
matrices ($\si_0$ is the unit matrix). However, the general
conclusion does not rely on the parameterization of the Dirac
matrices. Decomposing $h_\v k$ as $h_\v k=\sum_\mu \xi_{\v
k,\mu}\Ga_\mu$, one sees that the bulk dispersion is given by
$\w=\pm E_\v k$ with $E_\v k=\sqrt{\sum_\mu \xi_{\v k,\mu}^2}$. We
define $\Del=\min{E_\v k}$ as the bulk gap, which is assumed
nonzero. We specify to a layered hexagonal lattice illustrated in
Figs.1 (the impurities are considered in the next section), as
would be relevant to the
experiments.\cite{yazdanistm,kaptunikstm,xuestm,gomesstm} For
simplest purposes we consider a nearest-neighbor tight-binding
model and set $2t_n=1$ and $m=0$ henceforth. In this case
$\Del=1$.

\begin{figure}
\includegraphics[width=8.5cm]{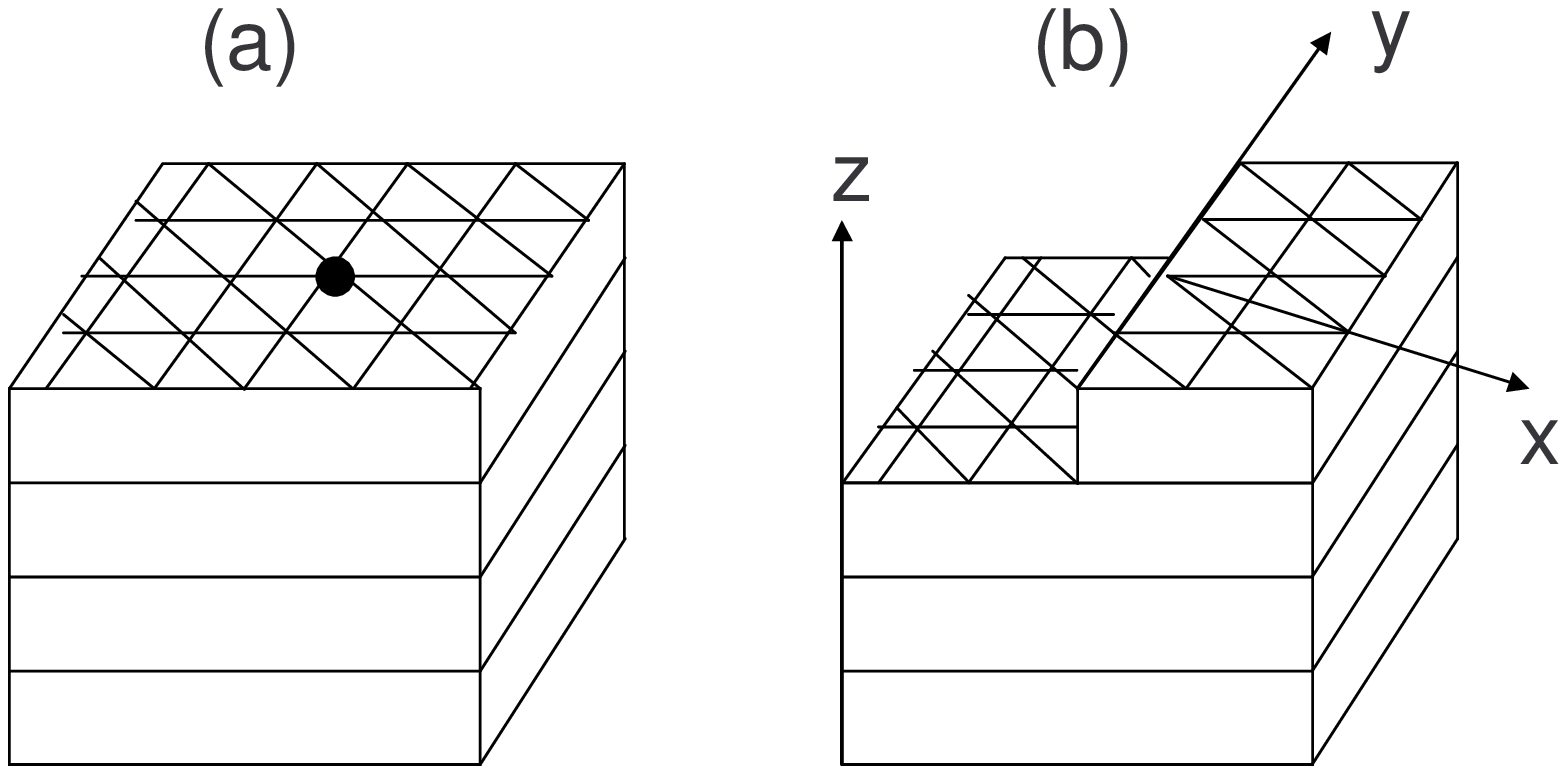}
\caption{Schematic plot of a topological insulator with layered
hexagonal structure, and a local impurity (filled circle) (a) and
(upper and lower) terraces (b) on the surface. Here $x$, $y$ and
$z$ denote the orthogonal axes, and the normal direction of the
terrace edge is along $x$. The inter-layer hopping is along the
vertical direction $z$.}
\end{figure}

We use a recursion method to obtain the surface Green's function.
In the presence of open surfaces normal to $z$, the planar
momentum $\v k_{||}=(k_x,k_y)$ is still a good quantum number. For
each $\v k_{||}$ the hamiltonian can be decomposed formally into
intra-layer part $\sum_n\psi_n^\dagger h^{2d}_{\v k_{||}}\psi_n$
and inter-layer part $\sum_n\psi_n^\dagger h_z\psi_{n+1}+{\rm
h.c.}$ where $n$ is the layer index, \eqa h^{2d}_{\v
k_{||}}=[m+2t_z-\sum_{\v b\neq \hat{z}}2t_b(\cos
k_b-1)]\Ga_0+\sum_{\v b\neq \hat{z}}2t_b\sin
k_b\Ga_b,\nonumber\eea and $h_z=-t_z(\Ga_0+i\Ga_3)$ before setting
our parametrization. Define $g=1/(z\v I-h^{2d}_{\v k_{||}})$ as
the Green's function for an isolated layer (here $z=\w+i0^+$ for
retarded Green's function, and $\v I$ is the $4\times 4$ unit
matrix), the surface Green's function for an N-layer lattice is
given by, recursively, \eqa G_N^{-1}=g^{-1}-h_z^\dagger G_{N-1}
h_z,\eea starting with $G_1=g$. Amazingly it is able to prepare
samples with a series of layers in recent
experiments.\cite{xueprivate} In this case the recursion method is
particularly useful to reveal the evolution of the surface states
as the sample thickness increases. However, in this paper, we are
only interested in infinite-layer lattices, for which
$G^{-1}=g^{-1}-h_z^\dagger G h_z$ holds for $G=G_\infty$.

\begin{figure}
\includegraphics[width=8.5cm]{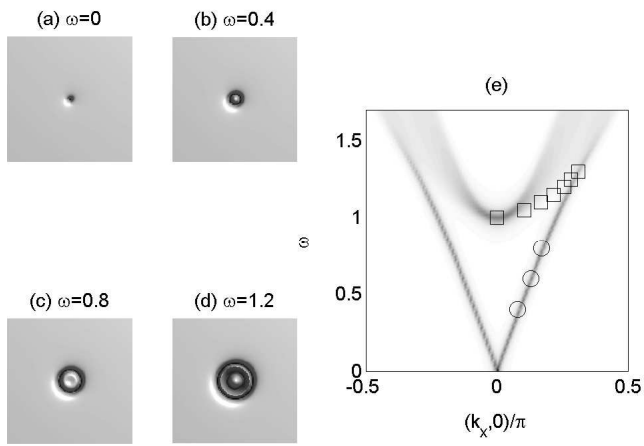}
\caption{(a)-(d): Intensity plot of $A(\v k_{||},\w)$ in the
planar $\v k_{||}$-space at $\w=0$, $0.4$, $0.8$, $1.2$. The
intensity scales with the darkness. The view field is bounded by
$2\pi\times 2\pi$ (within the hexagonal Brillouine zone). Camera
lighting is used to enhance the contrast. (e) Intensity plot of
$A(\v k_{||},\w)$ along a cut $(k_x,0)$ (gray). The open squares
and circles are extracted from LDOS oscillations in
Fig.\ref{figldoslinecut}(b) and Fig.\ref{figldosview},
respectively. \label{hexdispersion}}
\end{figure}

The spectral function for the surface states is given by \eqa A(\v
k_{||},\w)=-\frac{1}{\pi}\Im\Tr G_{\v k_{||}},\eea where the
planar momentum is indicated explicitly. The distribution in the
momentum space is plot in Figs.\ref{hexdispersion}(a)-(d) for a
series of $\w$ (only positive energies are displayed since the
spectrum is particle-hole symmetric). It is seen that there is
only a single dark spot at $\w=0$. This indicates that we have but
one Dirac cone in the Brillouine zone. The spot evolves to a ring
with increasing size for $\w<\Del=1$. For $\w>\Del$ bulk continuum
starts to contribute, forming doubled ring structure at $\w=1.2$
in combination with the Dirac ring. For still higher energies, the
ring structure is blurred (not shown). Since $A(\v k_{||},\w)$ is
quite rotationally symmetric in the $\v k_{||}$ space for the
energy window under concern (hexagonal structure shows up at
higher energies), we plot it in Fig.\ref{hexdispersion}(e) (gray)
along a line cut $(k_x,0)$. A clear massless Dirac dispersion is
seen. The bulk continuum starts at $\w=\Del$ and $\v k_{||}=0$,
and the outer border expands with increasing energies, while the
Dirac dispersion persists for $\w>\Del$ but eventually diminishes
for $\w>1.3$. The contributions from surface Dirac fermions and
bulk continuum are clearly separable. In comparison to pure Dirac
models, the surface Dirac fermions have a momentum dependent
spectral weight, the exact nature of which will be discussed
elsewhere.

\section{III. Terraces on the surface}

Experimentally terraces often appear on the surface. We assume the
upper terrace is one unit-cell higher than the lower one, as in
the experiment \cite{yazdanistm,kaptunikstm,xuestm,gomesstm} and
illustrated in Fig.1(b). The Green's function $\cG$ on the
terraces is obtained as follows. First, we map the hexagonal
lattice to a square lattice. The three principle translation axes
in the hexagonal lattice are mapped to the horizontal, vertical
and $45^o$ axes on the square lattice, for which we denote the
positions in real and momentum space as $(u,v)$ and $(p,q)$,
respectively. Consider the top two layers. We denote \eqa
G_q^{\al\bt}(u)=\frac{1}{M}\sum_p G^{\al\bt}(p,q)e^{ipu}\eea as
the $q$-resolved unperturbed Green's function, where $M$ is the
number of lattice sites along the $u$- or $v$-axis, and
$\al,\bt=U,L$ denotes the upper or lower layer. The Green's
function $G^{\al\bt}(p,q)$ can be obtained by coupling an isolated
upper layer (through $h_z$) to a surface already considered in the
previous section. Now setting $m\ra \infty$ for $u\leq 0$ on the
highest layer effectively depletes the left half of that layer,
leaving the lower-left (upper-right) layer mimic of the lower
(upper) terrace. The perturbed Green's function $\cG$ for the top
two layers can be obtained by the T-matrix
formalism,\cite{tmatrix} \eqa &&
\cG_q^{\al\bt}(u,u')=G_q^{\al\bt}(u-u')\nn && +\sum_{u_{a,b}\leq
0}G_q^{\al U}(u-u_a)T_q(u_a,u_b)G_q^{U\bt}(u_b-u'),\label{gq}\eea
where the layer index $U$ and the condition $u_{a,b}\leq 0$
indicate that the the depleted region is the upper-left layer, and
$T_q$ is the $q$-resolved T-matrix given by $T_q^{-1}(u_a,u_b)
=-G_q^{UU}(u_a-u_b)$. As what is required,
$\cG_q^{\al\bt}(u,u')=0$ whenever $u\al$ or $u'\bt$ falls on the
upper-left layer. We emphasize the Green's function obtained this
way contains all effects from the bulk below the terraces. The
LDOS on the terraces is given by, \eqa
\rho_\al(\w,u)=-\frac{1}{\pi M}\Im\Tr \sum_q
\cG_q^{\al\al}(u,u),\eea where $\al=L$ ($\al=U$) for $u<0$
($u>0$). Finally we map $u$ back to the normal displacement $x$
from the edge on the hexagonal lattice, as shown in Fig.1(b).

\begin{figure}
\includegraphics[width=8.5cm]{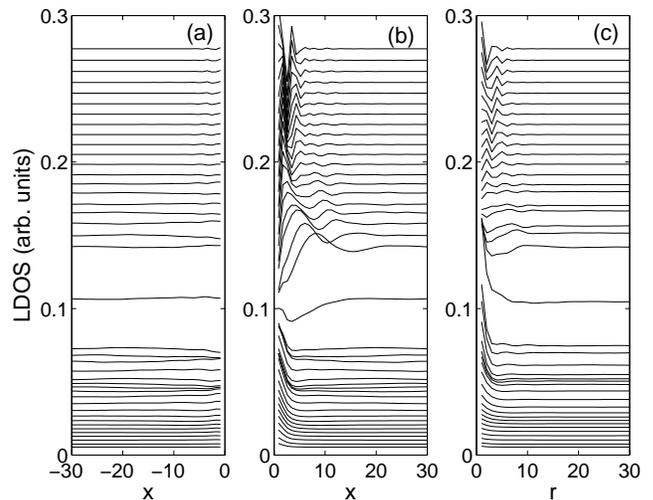}
\caption{LDOS as a function of the normal displacement $x$ on the
lower (a) and upper (b) terrace, and as a function of the radial
distance $r$ from a local impurity (c). In each panel the energy
increases from $\w=0$ at the bottom to $\w=2$ at the top, with
energy spacing $\Del\w=0.05$. The arrows indicate the lines at
$\w=1$ which coincides with the bulk gap.\label{figldoslinecut}}
\end{figure}

The LDOS as a function of $x$ is shown in
Fig.\ref{figldoslinecut}(a) and (b) for the lower and upper
terraces, respectively. In each panel, the energy increases from
$\w=0$ at the bottom up to $\w=2$ on the top, with energy interval
$\Del\w=0.05$. The arrow highlights the line with $\w=\Del$. For
reasons to be clarified, we defined the origin of $x$ differently
for the two terraces. We observe that the oscillation of LDOS on
the lower terrace is much weaker than that on the upper terrace.
This important asymmetry can be checked experimentally, and can be
understood as follows. As far as the LDOS (or the local Green's
function) is concerned, the upper terrace can be obtained by
setting $m\ra\infty$ on the left half of a surface. Therefore the
upper terrace effectively experience a hard wall on the left. On
the other hand, the lower terrace can be obtained by coupling a
surface to an isolated upper-right layer through $h_z$. Thus the
lower terrace effectively experiences a soft boundary on the
right. The asymmetry clearly follows from the difference in the
effective scattering mechanism, and makes it more sensible to
define the origin of $x$ as the boundary of such scattering in
Figs.\ref{figldoslinecut}(a) and (b).

Let us concentrate on Fig.\ref{figldoslinecut}(b). Here the LDOS
oscillation is negligibly small for $\w<\Del$, while more and more
peaks appear significant for $\w>\Del$. The LDOS oscillation is a
manifestation of quasiparticle scattering interference,\cite{qpi}
loosely referred to as the Friedel oscillation.\cite{friedel}  The
wavelength $\la_\w$, as is easily extracted from the first peak,
should correspond to the wave number $2k_\w$, the characteristic
momentum transfer during elastic quasiparticle scattering.
Therefore $k_\w=\pi/\la_\w$. By this means we obtain a dispersion
$\w$ vs. $k_\w$, which we plot in Fig.\ref{hexdispersion}(e) (open
squares). It clearly traces the outer border of the bulk
continuum, and is therefore not related to the surface Dirac
fermions. The lack of visible oscillations for $\w<\Del$ is in
nice agreement with the
experiment.\cite{yazdanistm,kaptunikstm,gomesstm} The oscillation
turns out to be more visible around a local impurity which we
discuss in the next section.

\section{IV. A local impurity on the surface}

\begin{figure}
\includegraphics[width=8.5cm]{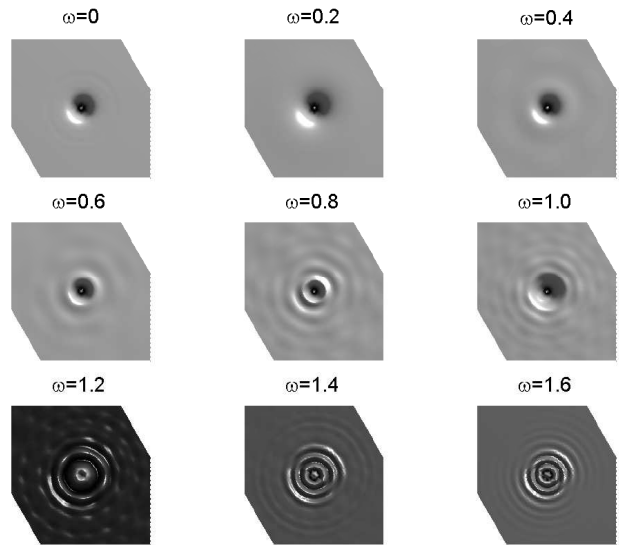}
\caption{Two-dimensional maps of the DOS near a local unitary
impurity for specific values of $\w$. The view field is bounded by
$26\times 26$. The strength scales with the darkness. The DOS at
the origin vanishes in our case, but it is artificially reset to
the value far from the impurity in order to enhance the contrast.
Camera lighting is used. \label{figldosview}}
\end{figure}

We consider a scalar impurity at the origin ($\v r=0$) on the
surface with the potential matrix $V\v I$, as illustrated in
Fig.1(a). (Non-scalar impurities \cite{magneticimpurity} can be
discussed along the same line but we assume scalar ones are more
likely to occur.) Given the unperturbed surface Green's function
$G$, the perturbed Green's function $\cG$ (in real space on the
surface) can be obtained again by the T-matrix formalism, \eqa
\cG(\v r,\v r')=G(\v r-\v r')+G(\v r)T G(-\v r'),\eea where \eqa
G(\v r)=\frac{1}{M^2}\sum_{\v k_{||}} G_{\v k_{||}} e^{i\v
k_{||}\cdot \v r},\eea and $T$ is given by $T^{-1}=V^{-1}\v
I-G(0)$. The LDOS is given by \eqa \rho(\w,\v
r)=-\frac{1}{\pi}\Im\Tr\cG(\v r,\v r).\eea We set $V\ra \infty$
for a unitary impurity. The LDOS along a principle translation
axis is plot in Fig.\ref{figldoslinecut}(c). We see LDOS
oscillation for $\w>\Del$ similar to that in (b), although it is
slightly weaker and has more complicated patterns. The oscillation
below $\Del$ is still weak but the peaks are visible for $\w>0.5$
and they shift toward the origin with increasing energy. To have a
better idea of the wavelength, we plot the LDOS map in
Figs.\ref{figldosview} for a few values of $\w$. The view field
bounded by $26\times 26$ is extracted from a $400\times 400$
surface. The oscillation, even if it is present, is beyond the
view field for $\w<0.4$, while it is clear (from the dark rings)
for higher energies. We extract the wave number $2k_\w$ from the
radii of the rings and plot $\w$ vs. $k_\w$ in
Fig.\ref{hexdispersion}(e) (open circles) for $\w<\Del$. It
clearly follows the Dirac dispersion. For $\w>\Del$, the
oscillation pattern is more complicated. In particular there are
supermodulations for $\w=1.2$ and $\w=1.4$, which can be
understood from the double-ring structure in the corresponding map
plot of the spectral function in Fig.\ref{hexdispersion}(d).

\section{V. Analytical results for Pure Dirac Models}

{\em 1D case}: For illustrative purpose and to set up notations,
we start with the single-particle hamiltonian in a 1D Dirac model,
in the momentum space, $h=k\si_n$, where $\si_n$ is one of the
three Pauli matrices and the Dirac velocity is set to unity
henceforth. The unperturbed Matsubara Green's function in the real
space is given by \eqa
G(x)&&=-\int\frac{dk}{2\pi}\frac{i\w_n\si_0+k\si_n}{\w_n^2+k^2}e^{ikx}
\nn  &&=-i\left[\w_n\si_0+|\w_n|\sign(x)\right]\phi(x),\eea where
$\si_0$ is the unitary matrix, $\w_n$ is the Matsubara frequency,
and we defined a kernel function \eqa \phi(x)=\int
\frac{dk}{2\pi}\frac{e^{ikx}}{\w_n^2+k^2}=
\frac{1}{2|\w_n|}e^{-|\w_n x|}.\eea  The on-site Green's function
$G(0)=-i\w_n\phi(0)\si_0$ turns out to be a scalar. Suppose there
is a local impurity potential $V\si_0$ at the origin, the
perturbed Green's function is conveniently obtained by the
T-matrix formalism, \eqa \cG(x,x')=G(0)+G(x)TG(-x'),\eea where
$T^{-1}=V^{-1}\si_0-G(0)\propto \si_0$ is the inverse of the
T-matrix. We will concentrate on $\del g(x)$, the trace of the
change of the on-site Green's function as a function of $x$. Upon
analytical continuation $i\w_n\ra \w+i0^+$, its imaginary part
gives the change of the LDOS. For $x\neq 0$, we find \eqa \del
g(x)&&=\Tr[G(x)TG(-x)]\nn && \propto \Tr
[(i\w_n\si_0+i|\w_n|\si_n)(i\w_n\si_0-i|\w_n|\si_n)]\nn &&\equiv
0.\eea Therefore right away from the impurity site, the LDOS is
unaffected at all. Moreover, using the above T-matrix formalism it
is easy to verify that the off-site Green's function
$\cG(x,x')=-G(x-x')$ for $x x'<0$ in the limit $V\ra \infty$. This
signifies perfect transmission (albeit with a phase lag of $\pi$),
a manifestation of the Klein Paradox. The mechanism behind this
effect is the chirality (the alignment between the momentum and
spin polarization) of the unperturbed eigen states. If an incoming
state were scattered backward, energy conservation requires a
flipping of the spin. But a scalar impurity can not flip the spin,
so the scattering matrix element between these states vanishes
identically.

For comparison, the retarded Green's function in 1D metal reads
$G(x)\sim -ie^{ik_\w|x|}/v_\w$ ($k_\w$ and $v_\w$ are the on-shell
momentum and group velocity at real energy $\w$). The oscillation
can be worked out by the T-matrix approach again. It has a wave
number $2k_\w$ and does not decay at all in the clean limit
implicitly assumed. (In reality, the oscillation must decay beyond
the mean free path.)

\begin{figure}
\includegraphics[width=8.5cm]{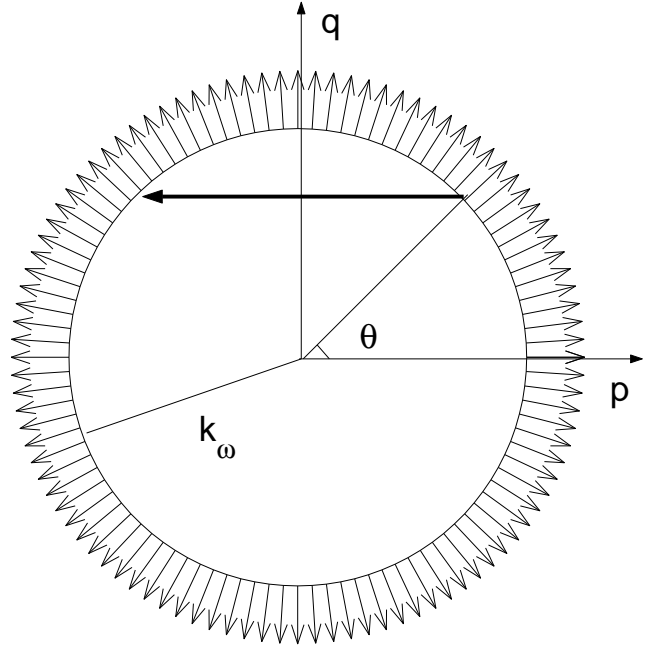}
\caption{Schematic plot of an energy shell at $\w$ with an
on-shell momentum $k_\w$. The radial arrows indicate the spin
polarizations. The long thick arrow indicates a scattering process
that conserves $q$ in the case of a domainwall with its normal
direction along the $p$-axis, and $\theta$ is the incident angle.
\label{figscatter} }
\end{figure}

{\em 2D case}: We write $h=\v k\cdot \si=p\si_x+q\si_y$ for 2D
Dirac fermions. We first consider a domain wall with scalar
potential $V\si_0$ along the $y$ direction. Then $q$ is still a
good quantum number. The scattering geometry is illustrated in
Fig.\ref{figscatter}, where the circle of radius $k_\w$ is an
energy shell, radial arrows indicate the momentum dependent spin
polarization and the long thick arrow indicate a scattering
process with incident angle $\theta$ (so that $q=k_\w\sin\theta$).
The $q$-resolved unperturbed Green's function is given by \eqa &&
G_q(x)=-\int\frac{dk}{2\pi}\frac{i\w_n\si_0+q\si_y+p\si_n}{\w_n^2+p^2+q^2}e^{ipx}
\nn &&=-\left[i\w_n\si_0+q\si_y+i\sqrt{q^2+\w_n^2}\si_x
\sign(x)\right]\phi_q(x),\eea where we defined \eqa
\phi_q(x)=\frac{1}{2\sqrt{q^2+\w_n^2}}e^{-\sqrt{q^2+\w_n^2}|x|}.\eea
In particular, \eqa G_q(0)=-(i\w_n\si_0+q\si_y)\phi_q(0),\eea is
no longer a scalar unless $q=0$. The $q$-resolved perturbed
Green's function is given by \cite{tmatrix} \eqa
\cG_q(x,x')=G_q(x-x')+G_q(x)T_qG_q(-x'),\eea with \eqa
T_q^{-1}=V^{-1}\si_0-G_q(0).\eea By straightforward algebra, we
find \eqa \del g_q(x)&&=\Tr\left[G_q(x)T_qG_q(-x)\right]\nn
&&=\frac{4q^2V_q^{-1}}{(i\w_n+V_q^{-1})^2-q^2}\phi_q(2x),\eea
where we defined $V_q=V\phi_q(0)=V/2\sqrt{q^2+\w_n^2}$.

Clearly the factor of $q^2$ in $\del g_q(x)$ follows from the
requirement of spin overlap $\sin^2\theta$ during elastic
scattering, as illustrated in Fig.(\ref{figscatter}). This leads
to $\del g_q(x)=0$ for $q=0$ (normal incidence), recovering the 1D
Klein paradox, as would have been anticipated. More importantly,
$\del g_q(x)\ra 0$ in the unitary limit $V\ra \infty$ (a hard
wall) for all values of $q$, and consequently in the real space
$\del g(x)=\int\del g_q(x)dq/2\pi\ra 0$ as long as $|x|>0$. This
explains nicely the numerical results in Sec.III for the terrace
that the LDOS barely oscillates below the bulk gap energy. (In
Sec.III we considered $m\ra \infty$, the effect of which is
however identical to $V\ra \infty$ here.) In the mean time, the
off-site Green's function $\cG_q(x,x')=-G_q(x-x')$ for $xx'<0$
holds, signaling perfect transmission in analogue to the 1D case.
This surprising result indicates that the Klein paradox also works
in 2D but only for a hard wall.

For general strength of $V$ (or soft walls), the asymptotic
behavior in the limit of $|\w_n x|\gg 1$ (where a saddle point
approximation is valid) is given by \eqa \del g(x)
&&=\int\frac{dq}{2\pi}\del g_q(x)\nn &&\sim
\frac{V}{(2+iV\sign{\w_n})^2\pi|x|}\sqrt{\frac{\pi}{|\w_n
x|}}e^{-2|\w_nx|},\eea where $q=k_\w\sin\theta$ is used. Upon
analytical continuation $|\w_n|\ra i\w$, this amounts to an
oscillation of LDOS with an energy-dependent wave number
$2k_\w=2|\w|$ and an envelope function that decays as
$|x|^{-3/2}$.

For comparison, in a 2D metal, $G_q(x)\sim
-ie^{ik_\w|x|\cos\theta}/v_\w\cos\theta$ in real frequency. By the
T-matrix formalism the domain wall leads to $\del g(x)\propto
-\int d\theta
e^{2ik_\w|x|\cos\theta}/(V^{-1}v_\w\cos\theta+i)\propto
e^{2ik_\w|x|}/\sqrt{k_\w |x|}$ for $k_\w|x|\gg 1$. So the LDOS
oscillation decays as $|k_\w x|^{-1/2}$ near a domain wall in a 2D
metal.\cite{metal} Notice that $k_{\w=0}$ is just the Fermi wave
number.

For a point impurity in 2D, the unperturbed Green's function for
Dirac fermions is given by \eqa G(\v
r)&&=-\int\frac{d^2k}{(2\pi)^2}\frac{i\w_n\si_0+\v
k\cdot\si}{\w_n^2+k^2}e^{i\v k\cdot \v r} \nn
&&=-i\left[\w_n\si_0+(|\w_n|+\frac{1}{2r})\hat{\v
r}\cdot\si\right]\phi(r),\eea where we find \eqa \phi(r)\sim
\frac{e^{-|\w_n|r}}{4\pi}\sqrt{\frac{2\pi}{|\w_n|r}}\eea in the
asymptotic limit $|\w_n|r\gg 1$. For the on-site Green's function,
a cutoff $\Lambda$ in momentum has to be introduced so that \eqa
G(0)= \frac{-i\w_n\si_0}{4\pi}\ln\frac{\Lambda^2}{\w_n^2}.\eea
Using the T-matrix formalism with $T^{-1}=V^{-1}\si_0-G(0)$, we
find \eqa \del g(\v r)&&=\Tr G(\v r)TG(-\v r)\nn &&\sim
\frac{1}{4\pi
V^{-1}+i\w_n\ln(\Lambda^2/\w_n^2)}\frac{e^{-2|\w_n|r}}{r^2},\eea
to leading order in $1/|\w_n|r$. Upon analytical continuation, we
see that the change of LDOS oscillates with an energy dependent
wave number $2k_\w=2|\w|$ and decays as $r^{-2}$. Notice that the
oscillation exists even for $V\ra\infty$, in contrast to the case
of the domain wall. This is consistent with the results in Sec.IV.
However, there is an important difference. Here a sharp resonance
state appears at $\w=0$ (in the unitary limit $V\ra\infty$) as
seen from analytical continuation of $\del g(\v r)$, a well known
result in many contexts.\cite{resonance} This resonance state is
absent in Sec.IV, as can be seen in Fig.\ref{figldoslinecut}(b) or
Figs.\ref{figldosview}. Qualitatively this is due to the fact that
surface Dirac fermions are not completely confined on the surface.
Depending on the planar momentum the wave function has varying
extent of amplitude on the surface. Therefore the impurity is only
partially seen by the surface Dirac fermions. Moreover, they exist
only in a limited regime in the momentum space. An accurate
discussion of these issues are left for future studies.

For comparison, in a 2D metal, $G(\v r)=-i\int d\theta e^{ik_\w
r\cos\theta}k_w/2\pi v_\w\propto e^{ik_wr}(k_\w r)^{-1/2}$ for
$k_\w r\gg 1$. By the T-matrix approach a local impurity leads to
$\del g(\v r)\propto G(\v r)^2\propto e^{2ik_\w r}/k_\w r$ for
$k_\w r\gg 1$. So the LDOS oscillation decays as $r^{-1}$ near a
local impurity in a 2D metal.\cite{metal}

Similar analysis could be proceeded for 3D Dirac fermions in the
presence of 2-, 1- and 0-dimensional impurities. As the impurity
states would not be easily probed by STM, we do not go into
detailed analysis. Some interesting behaviors are as follows.
First, a 2D hard wall does not lead to any oscillation in LDOS
away from the wall. For lower-dimension or general scalar
impurities the oscillation is present, and decays faster than that
for normal metals by a factor of $1/r$.

\section{VI. Summary and remarks}

We analyze the behavior of surface states of topological
insulators against local and extended impurities. Using an
effective 3D lattice model we first obtain the spectral function
of the surface states. This enables us to identify the
contributions from massless Dirac fermions and bulk continuum. We
then study the effects of a local impurity and a terrace on the
surface. Away from a local impurity, the spatial oscillation in
LDOS below the bulk gap $\Del$ is much weaker than that above
$\Del$. The LDOS oscillation is globally weak on the lower
terrace. While on the upper terrace, the LDOS oscillation is
barely visible below $\Del$, in perfect agreement with the STM
measurement, but it is significant above $\Del$. The asymmetry of
the LDOS oscillation on the lower and upper terraces is attributed
to the difference in the effective scattering mechanism. From the
LDOS oscillations we extract the modulation wave number $2k_\w$ as
a function of energy $\w$. The dispersion ($\w$ vs. $k_\w$)
follows that of Dirac fermions for $\w<\Del$, but it follows the
border of the bulk continuum for $\w>\Del$. The numerical results
combine to reveal that the surface Dirac fermions are rather
immune to the imperfections. We discuss such a behavior
analytically by pure Dirac models. Because of a cancellation due
to the alignment of momentum and spin polarization, the $2k_\w$
oscillation in LDOS, if present, decays asymptotically faster by a
factor of $1/r$ than that for usual fermions. Specifically, the
oscillation is absent for a hard domain wall. Such behaviors are
consistent with the 3d lattice model. We also find and discuss
differences between surface and pure Dirac fermions.

Before closing, some remarks are in order. First, our simple model
does not yield wavy energy shells (in the momentum space) for
surface Dirac fermions due to warping terms, which would arise if
longer-range hopping are considered. The warping term produces
partial nesting and is responsible for fine features in
Bi$_2$Te$_3$. However, the merit of our simple model is it
uncovers material-independent qualitative difference between
surface Dirac fermions and bulk continuum. A material-dependent
calculation will be provided elsewhere. Second, we notice that a
real terrace edge is not likely a straight line. This may be
mapped to a straight edge but with excess impurities nearby.
According to our results of both local impurity and terrace, a
wiggling terrace edge should also lead to visible oscillations
below the bulk gap, consistent with experiments.\cite{xuestm}
Finally, in a model with multiple Dirac cones for the surface
states, elastic scattering between cones is not suppressed by the
chirality, and is therefore expected to induce visible LDOS
oscillations even below the bulk gap.

While finalizing the writing of this work, we become aware of a
related work using a 2D continuum model with warping
terms.\cite{hujp}

The work in Nanjing was supported by NSFC 10974086 and 10734120,
the Ministry of Science and Technology of China (under the Grant
No. 2006CB921802 and 2006CB601002) and the 111 Project (under the
Grant No. B07026). The work in Hong Kong was supported by the RGC
grand of Hong Kong.

\end{document}